\documentclass[aps,prb,twocolumn,english,superscriptaddress,citeautoscript,preprintnumbers,amsmath,amssymb,floatfix,footinbib]{revtex4-2}
\usepackage{graphicx}
\usepackage{graphics}               
\usepackage{subfigure}
\usepackage{bm}   
\usepackage{amssymb } 
\usepackage{array}       
\usepackage[breaklinks=true,colorlinks,citecolor=blue,linkcolor=blue,urlcolor=blue]{hyperref}
\usepackage{float}
\usepackage[table]{xcolor}
\usepackage{cancel}
\usepackage[normalem]{ulem}

\definecolor{Cyan}{rgb}{0.4,1,0.7}
\definecolor{LightCyan}{rgb}{0.8,1,0.85}

\begin{document}
	
\title{Large non-saturating Nernst thermopower and magnetoresistance in compensated semimetal ScSb}

\author{Antu Laha}\email[]{antulaha.physics@gmail.com}
    \affiliation{Department of Physics and Astronomy, Stony Brook University, Stony Brook, New York 11794-3800, USA}
    \affiliation{Condensed Matter Physics and Materials Science Division, Brookhaven National Laboratory, Upton, New York 11973-5000, USA}
 
 \author{Sarah Paone}
    \affiliation{Department of Physics and Astronomy, Stony Brook University, Stony Brook, New York 11794-3800, USA}
    \affiliation{Condensed Matter Physics and Materials Science Division, Brookhaven National Laboratory, Upton, New York 11973-5000, USA}

\author{Niraj Aryal }
    \affiliation{Condensed Matter Physics and Materials Science Division, Brookhaven National Laboratory, Upton, New York 11973-5000, USA}    
    
\author{Qiang Li}\email[]{qiang.li@stonybrook.edu}
    \affiliation{Department of Physics and Astronomy, Stony Brook University, Stony Brook, New York 11794-3800, USA}
     \affiliation{Condensed Matter Physics and Materials Science Division, Brookhaven National Laboratory, Upton, New York 11973-5000, USA}

\begin{abstract}
Today, high-performance thermoelectric and thermomagnetic materials operating in the low-temperature regime, particularly below the boiling point of liquid nitrogen remain scarce. Most thermomagnetic materials reported to date exhibit a strong Nernst signal along specific crystallographic directions in their single-crystal form. However, their performance typically degrades significantly in the polycrystalline form. Here, we report an improved Nernst thermopower of $\sim$ 128 $\mu$V/K at 30 K and 14 T in polycrystalline compensated semimetal ScSb, in comparison to that was observed in single crystal ScSb previously. The magnetic field dependence of Nernst thermopower shows a linear and non-saturating behavior up to 14 T. The maximum Nernst power factor reaches to $\sim 240 \times 10^{-4}$ W m$^{-1}$ K$^{-2}$ and Nernst figure of merit reaches to $\sim 11 \times 10^{-4}$ K$^{-1}$. Polycrystalline ScSb also shows a large non-saturating magnetoresistance of $\sim 940 \%$ at 2 K and 14 T. These enhanced properties originate from better electron–hole compensation, as revealed by Hall resistivity measurements. The cubic symmetry and absence of anisotropy in ScSb allow its polycrystalline form to achieve similar enhanced thermomagnetic and electromagnetic performance  comparable to that of the single crystal.
\end{abstract}

\maketitle

\section{Introduction}
Thermoelectric systems enable the direct conversion of heat into electricity and vice versa, positioning them as promising technologies for applications like harvesting energy from waste heat and achieving solid-state cooling \cite{Bell_Science_2008, Funahashi_book, Snyder_Nat_Matr_2008, Mao_Nat_Matr_2021}. The efficiency of these devices in generating power is largely influenced by the power factor (PF), calculated using the formula $PF = \sigma ~S^2$, where $\sigma$ is the electrical conductivity and $S$ is the thermopower. Depending on how the voltage output aligns with the applied temperature gradient, two primary effects are observed: the longitudinal Seebeck effect and the transverse Nernst effect. In the Seebeck effect, the voltage is generated along the same direction as the thermal gradient. In contrast, the Nernst effect is activated in the presence of a magnetic field—producing a voltage that is perpendicular to the temperature difference. Accordingly, the power factors for these effects are represented as $PF_S = \sigma_{yy}~S_{xx}^2$ for Seebeck and $PF_N = \sigma_{yy} ~ S_{yx}^2$ for Nernst, with $S_{xx}$ and $S_{yx}$ denoting the Seebeck and Nernst thermopower respectively. Nernst-based thermoelectric devices offer a simpler design compared to Seebeck-based devices, as they eliminate the need for paired p-type and n-type components. This not only streamlines the fabrication process but also minimizes both thermal and electrical resistance in the final device.

Recently, topological semimetals such as Cd$_3$As$_2$, ZrTe$_5$, NbP, PtSn$_4$, Mg$_2$Pb, NbSb$_2$, and WTe$_2$ have emerged as exceptional candidates for enhancing the Nernst effect \cite{Cd3As2_PRL_2017, ZrTe5_PRB_2021, PtSn4_Research_2020, Mg2Pb_natComn_2021, NbSb2_NatComn_2022, WTe2_NanoLett_2018, WTe2_NatComn_2022, Review_TM_AFM_2025}. The enhancement of the Nernst thermopower in these materials is believed to stem from the combined influence of three key mechanisms: (i) the phonon drag effect, which involves momentum transfer from lattice vibrations to charge carriers; (ii) the electron-hole compensation effect, which balances carrier populations to optimize transport; and (iii) the presence of topologically non-trivial band structures, which significantly alter the charge transport behavior in magnetic fields. Notably, these materials exhibit high thermomagnetic efficiency in their single-crystal form, with the Nernst thermopower being prominent along specific crystallographic axes. However, their performances reduced considerably in the polycrystalline form. For instance, NbP single crystals exhibit a large Nernst thermopower of 800$\mu$V/K, which drops to 90$\mu$V/K in polycrystalline samples \cite{NbP_poly_EEC_2018}. Polycrystalline samples are generally more suitable for device applications due to their ease of synthesis in large bulk size, whereas single crystals are limited in size by their growth methods. The reduction of the Nernst thermopower in polycrystalline materials depends on their crystal structure, as the measured value represents an average of the Nernst responses along different crystallographic directions. Materials that crystallize in a cubic structure may serve as more promising candidates, owing to the absence of crystallographic anisotropy.

The MX family of compounds (M = Sc, Y, La, Ce, Pr, Nd, Sm, Gd, Tb, Yb, Lu; X = As, Sb, Bi) crystallizes in a cubic structure and their single crystals have been extensively studied for exceptionally large non-saturating magnetoresistance driven by electron–hole compensation \cite{ScSb_PRB_2018, LaSb_NaturePhy_2015, LaSbBi_PRB_2018, YBi_PRB_2019, YBi_PRB_2021, YSb_SR_2016, YbAs_PRB_2020, PrSb_PRB_2017, Pr_Sm_Sb_Bi, NdSb_PRB_2016, NdBi_2025, GdBi_PRB_2023, Lu_Y_Bi_PRB_2018, TbSb_MTP_2022}. However, another intriguing phenomenon arising from electron–hole compensation - the large Nernst effect, has not yet been investigated in any member of this family except ScSb. Despite the topologically trivial electronic band structure, we recently reported a large Nernst thermopower in ScSb single crystals, due to electron-hole compensation effect. \cite{ScSb_arxiv_2025}. In the present work, we report both a pronounced non-saturating Nernst thermopower and a large non-saturating magnetoresistance in polycrystalline ScSb. The magnetoresistance is 940$\%$ at 2 K and 14 T. The Nernst thermopower reaches a peak value of 128 $\mu$V/K at 14 T and 30 K, resulting in a maximum Nernst power factor of 240 $\times 10^{-4}$ W m$^{-1}$ K$^{-2}$, which is nearly seven times the value found in the single crystal samples at the same temperature and magnetic field \cite{ScSb_arxiv_2025}. The Nernst power factor for polycrystalline ScSb is even larger than that of Seebeck power factor ($PF_S$) found in state-of-the-art thermoelectric materials such as Bi$_2$Te$_3$ ($PF_S$ = 42 $\times 10^{-4}$ W m$^{-1}$ K$^{-2}$) \cite{Bi2Te3_Science_2008}, PbTe ($PF_S$ = 25 $\times 10^{-4}$ W m$^{-1}$ K$^{-2}$) \cite{PbTe_Nature_2011}, and SnSe ($PF_S$ = 10 $\times 10^{-4}$ W m$^{-1}$ K$^{-2}$) \cite{SnSe_Nature_2014}. 

\section{Methods}
Polycrystalline ScSb samples were synthesized through a combined approach of solid-state reaction and spark plasma sintering (SPS). Stoichiometric amounts of Sc ingot (99.99$\%$), and Sb chips (99.999$\%$) were put together in an evacuated quartz ampoule, and then the ampoule was kept inside a muffle furnace at a temperature of 800$^\circ$C for 72 hours to achieve a complete reaction. The reacted product was ground into fine powders, and loaded into a graphite die with inner diameter of 15 mm. The content was immediately sintered at 850$^\circ$C for 15 min under an uniaxial pressure of 50 MPa using SPS. The relative density of the compacted pellet was $\sim 96\%$. 

The powders of polycrystalline ScSb were characterized by x-ray
diffraction (XRD) technique using Cu-K$_\alpha$ radiation in a Rigaku miniflex diffractometer to determine the crystal structure and the phase purity. The XRD data were analyzed by Rietveld structural refinement using the FULLPROF package \cite{Fullprof_1993}. The chemical compositions were confirmed by energy dispersive x-ray spectroscopy (EDS) measurements in a a JEOL JSM-7600F scanning electron microscope. Electronic transport measurements were carried out in a physical property measurement system (PPMS, Quantum Design) via the standard four-probe method. The Seebeck thermopower and Nernst thermopower were measured using four-probe technique on a standard thermal transport option (TTO) platform, and a modified one where the Cu wires for measuring voltage signals were separated from the Cernox 1050 thermometers \cite {NbSb2_NatComn_2022}.

\begin{figure}
	\centering
	\includegraphics[width=0.99\linewidth]{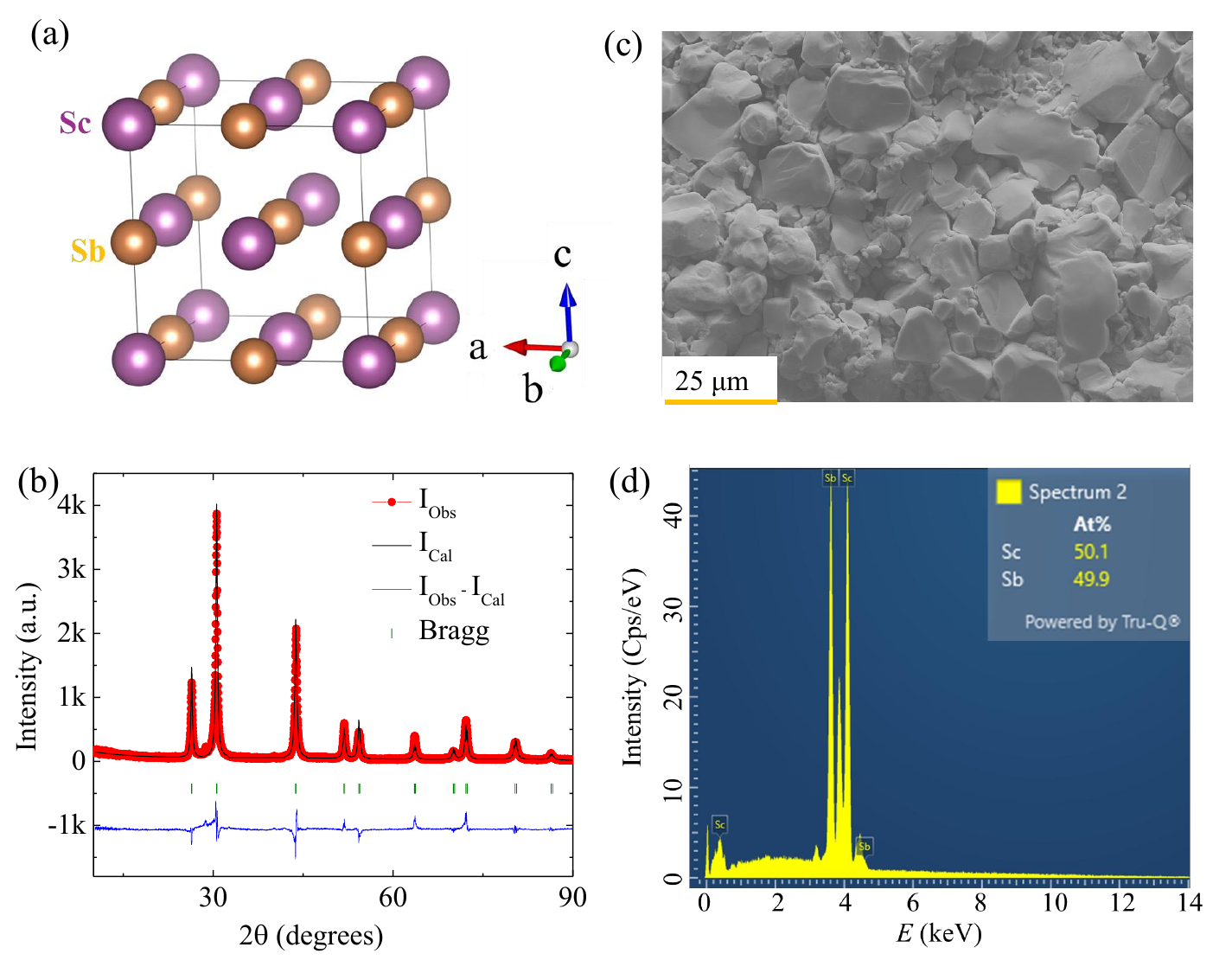} 
	\caption{(a) Cubic crystal structure of ScSb. (b) Refined powder x-ray diffraction pattern of polycrystalline ScSb, recorded at room temperature. The observed intensity (red scattered points), Rietveld refinement fit (solid black line), difference between the experimentally observed and calculated intensities (solid blue line), and Bragg peak positions (vertical green bars) are shown. (c) Low magnification scanning electron microscopy (SEM) image of fractured polycrystalline ScSb, showing the grains. (d) Energy-dispersive X-ray spectroscopy confirms the stoichiometry of the compound.}
	\label{Fig1}
\end{figure}

\section{Results and Discussions}
The cubic crystal structure (space group $F\bar m 3m$) of polycrystalline ScSb is confirmed by the Rietveld refinement of powder x-ray diffraction pattern [Fig.\ref{Fig1}(a) and \ref{Fig1}(b)]. The low magnification scanning electron microscopy (SEM) image as shown in Fig.\ref{Fig1}(c) indicates that the grain size of synthesized polycrystalline ScSb is in several microns. The EDS analysis confirms the single-phase nature and almost perfect stoichiometry of the polycrystalline sample [Fig.\ref{Fig1}(d)].

\begin{figure}
	\centering
	\includegraphics[width=1.0\linewidth]{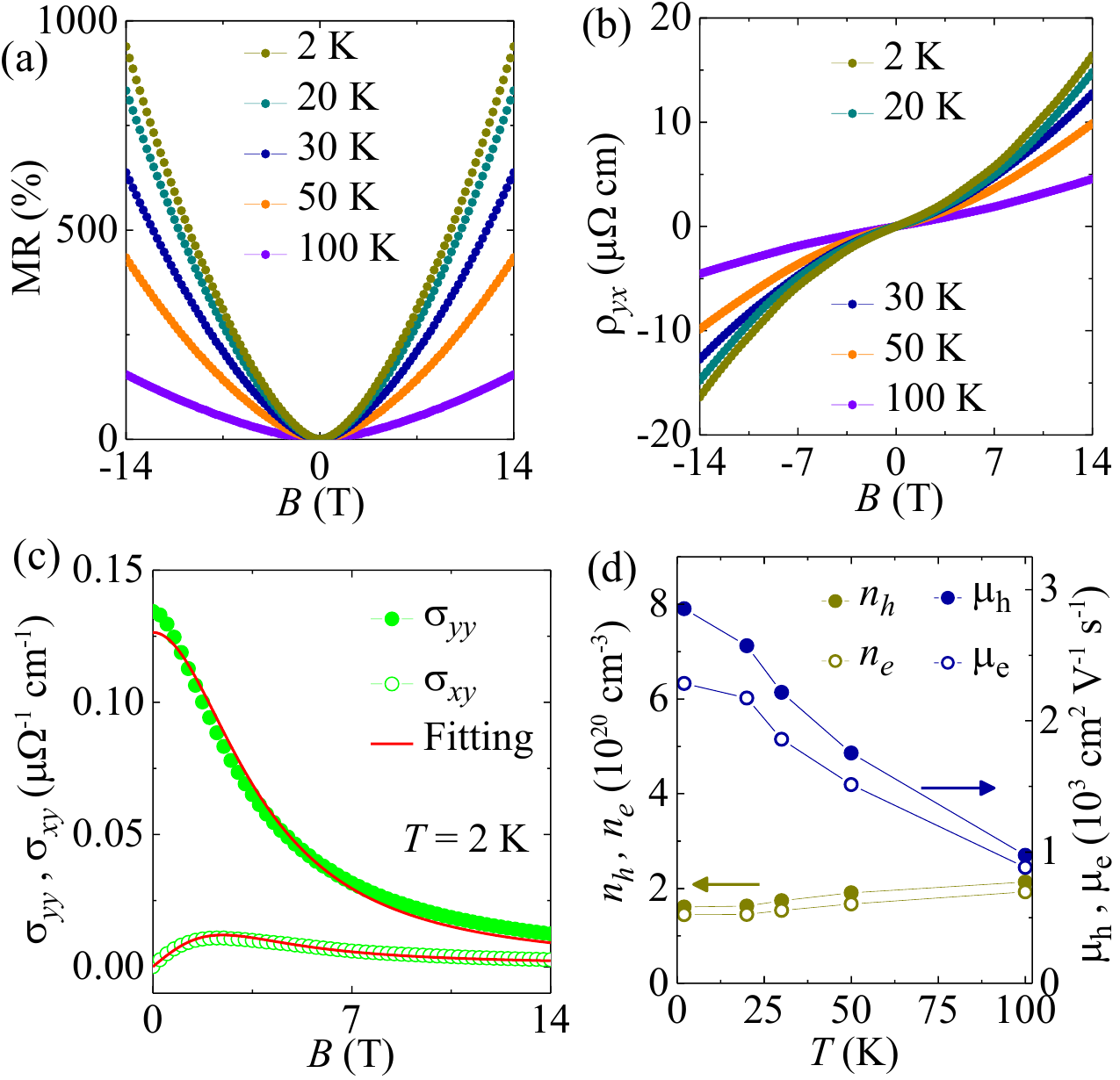} 
	\caption{(a) Longitudinal resistivity ($\rho_{yy}$), and (b) Hall resistivity ($\rho_{yx}$) as function of $B$ at various temperatures. (c) Simultaneous fitting of longitudinal conductivity ($\sigma_{yy}$) and Hall conductivity ($\sigma_{xy}$) with two carries model (equation (\ref{xx}) and (\ref{yx})). (d) Hole (electron) density $n_h$ ($n_e$) and mobility $\mu_h$ ($\mu_e$) as a function of temperature.}
	\label{Fig_MR}
\end{figure}

As shown in Fig.~\ref{Fig_MR}(a), polycrystalline ScSb exhibits a large and non-saturating magnetoresistance (MR) of $\sim940\%$ at 2 K and 14 T. Such a high MR is atypical for polycrystalline materials. On one hand grain boundaries generally act as scattering centers that impede the motion of charge carriers, on the other hand grain boundaries can disrupt the coherent cyclotron motion of electrons and holes in magnetic fields, thereby reducing the overall MR in some cases. This grain boundary scattering effect  in magnetic fields is small in conventional metal or semiconductors. However, in certain topological semimetals such as NbSb$_2$ and NbP, a large MR has been observed even in polycrystalline form \cite{NbP_poly_AFM_2022, NbSb2_poly_EES_2023}. This unusual behavior is often attributed to a combination of factors: high carrier mobility enabled by linearly dispersing electronic bands, topologically protected states that reduce backscattering, and nearly perfect compensation between electron and hole carriers. These mechanisms can collectively sustain high MR despite the presence of grain boundaries. In addition, silver chalcogenides such as Ag$_{2+\delta}$Se and Ag$_{2+\delta}$Te, have been reported to show substantial MR ($\sim 200\%$) in polycrystalline samples \cite{Ag2Se_Nature_1997}.

To investigate the origin of the large MR in ScSb, we carried out Hall resistivity ($\rho_{yx}$) measurements over a range of temperatures [see Fig.~\ref{Fig_MR}(b)]. The nonlinear field dependence of $\rho_{yx}$ is indicative of multi-carrier transport, with both electrons and holes contributing to conduction. To quantify their respective densities and mobilities, we consider a semiclassical two-carrier model to simultaneously fit the Hall conductivity ($\sigma_{xy}$) and longitudinal conductivity ($\sigma_{yy}$) as shown in Fig.\ref{Fig_MR}(c). According to semiclassical two-carrier model \cite{Two_Carrier_Model, TaAs_PRB_2017}, the $\sigma_{yy}$ and $\sigma_{xy}$ can be expressed as

\begin{equation}
  \sigma_{yy} = e \left[\frac{n_h \mu_h}{1 + \mu_h^2 B^2} + \frac{n_e \mu_e}{1 + \mu_e^2 B^2} \right]
  \label{xx}
\end{equation}

\begin{equation}
     \sigma_{xy} = e B \left[\frac{n_h \mu_h^2}{1 + \mu_h^2 B^2} - \frac{n_e \mu_e^2}{1 + \mu_e^2 B^2}\right]
     \label{yx}
\end{equation}

where $n_h$ ($n_e$) and $\mu_h$ ($\mu_e$) are the hole (electron) density and mobility, respectively. The $\sigma_{yy}$ and $\sigma_{xy}$ are obtained using the expressions $\sigma_{yy} = \frac{\rho_{yy}}{\rho_{yy}^2 + \rho_{yx}^2}$, and $\sigma_{xy} = \frac{\rho_{yx}}{\rho_{yy}^2 + \rho_{yx}^2}$. The calculated carrier densities (error bar: $\pm 5\%$) and mobilities (error bar: $\pm 5\%$) from the two carriers model fitting are plotted as a function of temperature in Fig.\ref{Fig_MR}(d). The low carrier density of $\sim 1.5 \times 10^{20}$ cm$^{-3}$ suggests a semimetallic nature of the polycrystalline sample. The carrier density ratio ($n_h/n_e$ is $\sim 1.1$) indicates the occurrence of electron-hole compensation. This, combined with the high carrier mobility ($\mu = \sqrt{\mu_h \mu_e}$) of $\sim$ 2553 cm$^2$V$^{-1}$s$^{-1}$ at 2 K, is primarily responsible for the large non-saturating magnetoresistance observed in polycrystalline ScSb.

\begin{figure}
	\centering
	\includegraphics[width=0.99\linewidth]{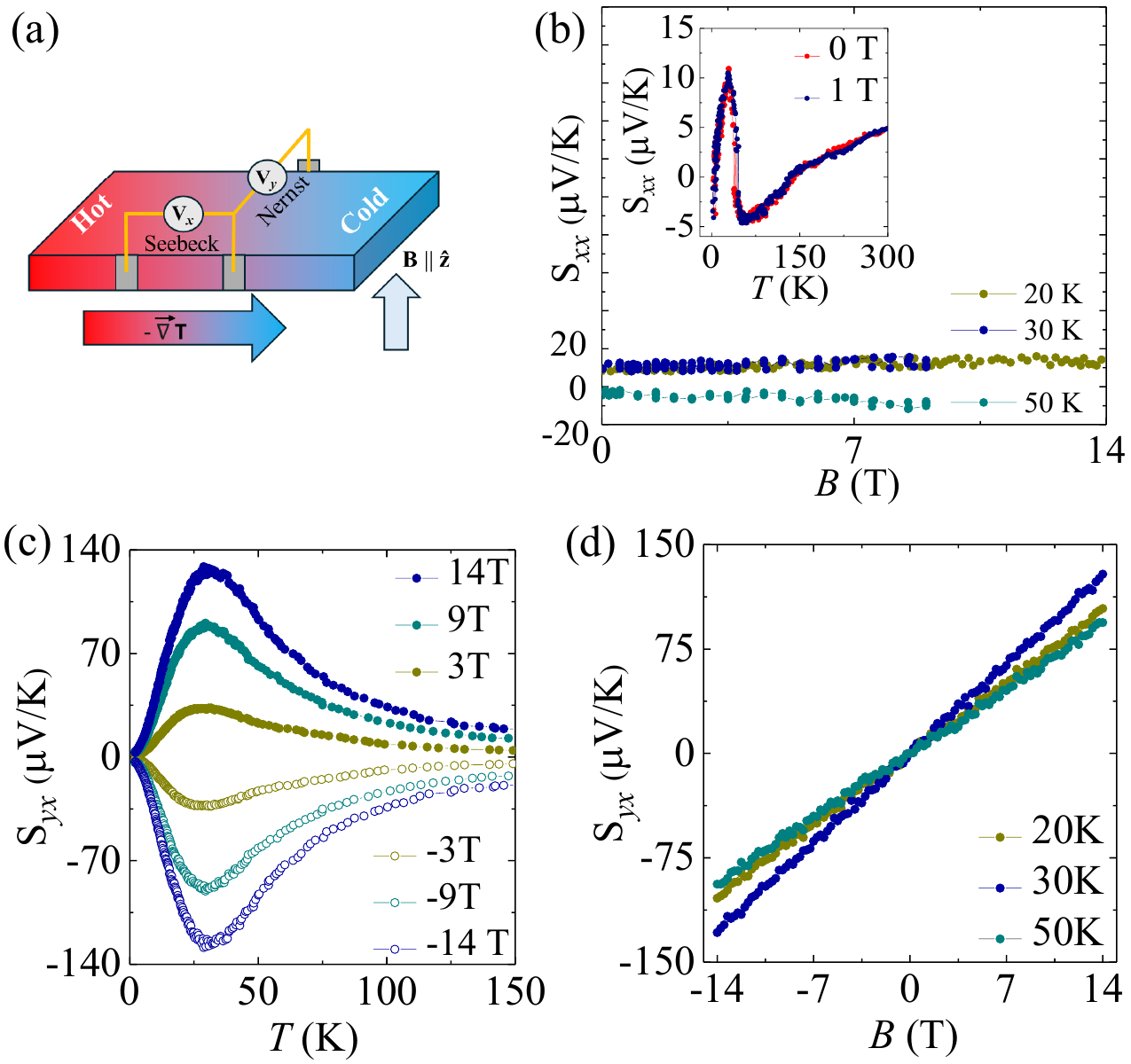} 
	\caption{(a) Schematic diagram for Seebeck and Nernst effect measurements. (b) Seebeck thermopower ($S_{xx}$) as a function of magnetic field. The inset shows temperature dependence of $S_{xx}$. (c) Nernst thermopower ($S_{yx}$) as a function of temperature at various magnetic fields. (d) Nernst thermopower as function of magnetic field at various temperatures.}
	\label{Fig_Sxy}
\end{figure}

\begin{table}
\begin{center}
\caption{\label{Parameters}Comparison of maximum Nernst thermopower ($S_{yx}$), hole-to-electron density ratio ($n_h/n_e$), and effective mobility ($\mu = \sqrt{\mu_h \mu_e}$) between single-crystal (ref. \cite{ScSb_arxiv_2025}) and polycrystalline sample (this work) of ScSb.}

\label{SC_Poly}
\begin{tabular}{p{65pt}p{80pt}p{30pt}p{55pt}}

\hline\hline

ScSb  & $S_{yx}$($\mu$V/K) & $n_h/n_e$ & $\mu$(cm$^2$V$^{-1}$s$^{-1}$) \\ 

 \hline 

Single-crystal & $\sim 47$ (12 K, 14 T) & 0.8   & 2210 (2 K) \\
 
Polycrystalline & $\sim 128$ (30 K, 14 T) & 1.1  & 2553 (2 K)\\
 
 \hline\hline
\end{tabular}
\end{center}
\end{table}

Seebeck thermopower ($S_{xx}$) and Nernst thermopower ($S_{yx}$) are measured as a function of temperature for polycrystalline ScSb as shown in Fig.\ref{Fig_Sxy}(b) and \ref{Fig_Sxy}(c). The $S_{xx}$ in polycrystalline ScSb is positive in the temperature range 135 K $-$ 300 K, shows a crossover from positive to negative at $\sim$ 135 K and another crossover from negative to positive at $\sim$ 42 K as shown in the inset of Fig.\ref{Fig_Sxy}(b). This behavior indicates that the sample is very close to electron-hole compensation. In contrast, single-crystal ScSb shows negative $S_{xx}$ between 100 K and 300 K, with a single crossover from negative to positive at $\sim$ 100 K \cite{ScSb_arxiv_2025}. The $S_{xx}$ in polycrystalline ScSb shows a strong peak of $\sim$ 10 $\mu$V/K at $\sim$ 30 K. In ScSb single crystal, a similar peak has been observed at 12 K with a value of $\sim$ 2.8 $\mu$V/K. A distinct peak in the longitudinal Seebeck coefficient ($S_{xx}$) is a well-known feature in many thermoelectric and thermomagnetic materials and is typically attributed to the phonon-drag mechanism. As temperature increases, phonons with higher momentum become thermally activated. When these long-wavelength acoustic phonons transfer their momentum efficiently to charge carriers near the Fermi surface, a drag effect arises, producing an enhancement in thermopower at low temperatures \cite{Goldsmid, Delves_1964, NbSb2_NatComn_2022, NbSb2_poly_EES_2023}. Notably, the temperature dependence of $S_{xx}$(T) at zero field and under an applied magnetic field of 1 T overlaps throughout the entire temperature range from 2 K to 300 K, as shown in the inset of Fig.\ref{Fig_Sxy}(b). Furthermore, we measured the magnetic field dependence of $S_{xx}$ at several temperatures as shown in Fig.\ref{Fig_Sxy}(b). The $S_{xx}$ remains nearly unchanged with increasing magnetic field.

Fig.~\ref{Fig_Sxy}(c) shows the transverse thermopower ($S_{yx}$), or Nernst signal, as a function of temperature at various magnetic fields. A pronounced peak appears around 30 K, similar to that observed in $S_{xx}$, which is attributed to the phonon drag effect. This peak becomes more significant under stronger magnetic fields, reaching $\sim$ 128 $\mu$V/K at 14 T. When the magnetic field is reversed to -14 T, $S_{yx}$ exhibits a similar trend but with an opposite sign, consistent with antisymmetric Nernst response. Large Nernst thermopower has previously been reported in several topological semimetals, where it is often attributed to the combined effects of non-trivial band topology and near-perfect electron-hole compensation. Interestingly, in the case of polycrystalline ScSb, a material with topologically trivial electronic bands, we also observe a significant enhancement of the Nernst effect, suggesting that electron-hole compensation alone can lead to substantial thermomagnetic response. The observed $S_{yx}$ in polycrystalline ScSb is larger than that of its single-crystalline counterpart ($S_{yx} \sim$ 35 $\mu$V/K at 12 K and 14 T) \cite{ScSb_arxiv_2025}, likely due to improved electron-hole compensation. A comparison of the Nernst thermopower, hole-to-electron density ratio, and carrier mobility between polycrystalline and single-crystalline ScSb is presented in table \ref{SC_Poly}. The peak value of $S_{yx}$ in polycrystalline ScSb is comparable to that of polycrystalline Mg$_3$Bi$_2$ (127 $\mu$V/K at 13.5 T and 15 K) \cite{Mg3Bi2_poly_AM_2022}, and even exceeds that seen in other polycrystalline thermomagnetic compounds such as NbP (90 $\mu$V/K at 9 T and 136 K) \cite{NbP_poly_EEC_2018}, Bi$_{77}$Sb$_{23}$ (88.5 $\mu$V/K at 6 T and 50 K) \cite{Bi77Sb23_APL_2020}, Te-doped Bi$_{77}$Sb$_{23}$ (45 K $\mu$V/K at 6 T and 200 K) \cite{Bi77Sb23_APL_2020} and Ag$_{2(1+x)}$Se (15 $\mu$V/K 5 T and 300 K) \cite{Ag2(1+x)Se_JSSc_2020}.
\begin{figure}
	\centering
	\includegraphics[width=0.95\linewidth]{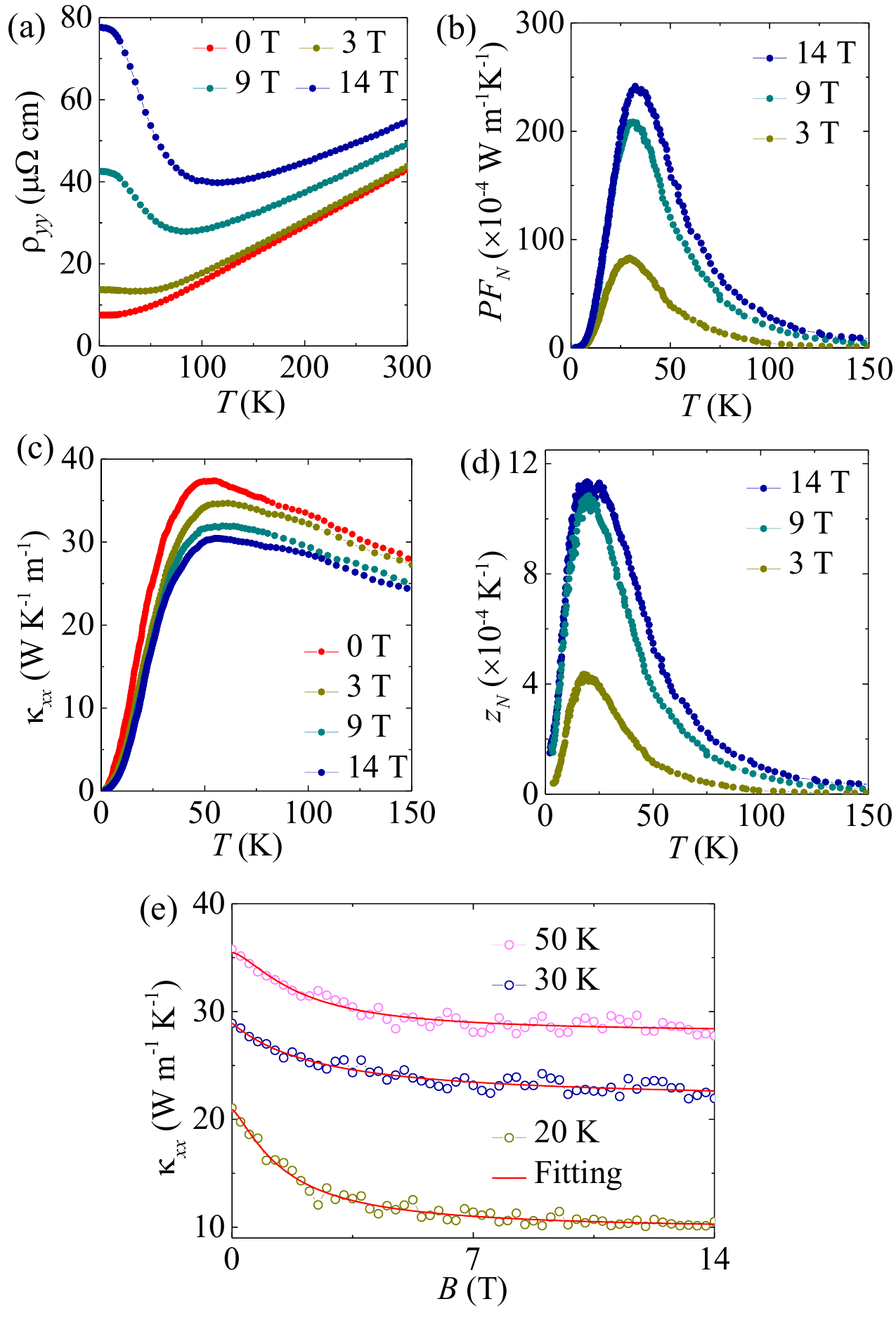} 
	\caption{(a) Longitudinal electrical resistivity ($\rho_{yy}$), (b) Longitudinal thermal conductivity ($\kappa_{xx}$), (c) Nernst power factor ($PF_N$), and (d) Nernst figure of merit ($z_N$) as a function of temperature. (e) Measured thermal
conductivity as a function of magnetic field at different temperatures. The symbols are experimental data and the lines are the fitting curves to the equation (\ref{kxx}) [fitting error : $\pm 10\%$].}

	\label{Fig_PFN}
\end{figure}

\begin{table}[t]
\begin{center}
\caption{\label{Parameters}Parameters used to fit the measured thermal conductivity of polycrystalline ScSb.}

\label{Parameters}
\begin{tabular}{p{25pt}p{55pt}p{55pt}p{40pt}p{20pt}}
\hline\hline
 $T$(K) & $\kappa_l$ (WK$^{-1}$m$^{-1}$)& $\kappa_e$ (WK$^{-1}$m$^{-1}$)& $\eta$ (T$^{-s}$)& $s$ \\ 
 \hline 
 
 20 & 9.88 & 12.88  & 0.58 & 1.32 \\
 
 30 & 20.87 & 8.39 & 0.39& 1.05 \\
 
 50 & 27.24 & 8.56 & 0.37 & 1.49 \\
 
 \hline\hline
\end{tabular}
\end{center}
\end{table}

\begin{figure*}
	\centering
	\includegraphics[width=0.8\linewidth]{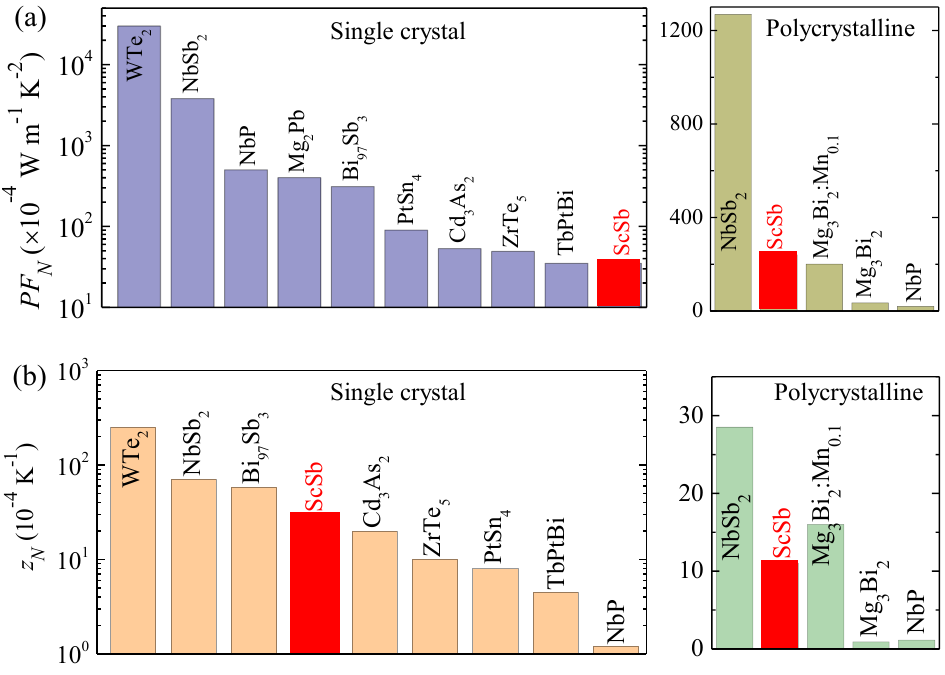} 
	\caption{Comparison of the maximum (a) Nernst power factor $PF_N$ and (b) Nernst figure of merit $z_N$ of ScSb and other thermomagnetic materials \cite{WTe2_NatComn_2022, NbSb2_NatComn_2022, NbP_JPCM_2017, NbP_PRB_2016, Mg2Pb_natComn_2021, PtSn4_Research_2020, Cd3As2_PRL_2017, ZrTe5_PRB_2021, TbPtBi_AM_2023, Bi97Sb3_PRB_2018, ScSb_arxiv_2025, Mg3Bi2_Science_2019, Mg3Bi2_poly_AM_2022, NbP_poly_AFM_2022, NbSb2_poly_EES_2023}.}
	\label{Fig_comp}
\end{figure*}

The $S_{yx}$ linearly increases with increasing magnetic fields, without showing a signature of saturation up to 14 T as shown in Fig.\ref{Fig_Sxy}(d). Such a non-saturating $S_{yx}$ is previously reported in topological semimetal such as NbSb$_2$ \cite{NbSb2_NatComn_2022, NbSb2_poly_EES_2023}, and it is highly uncommon in topologically trivial materials. In general, Nernst thermopower saturates at high magnetic fields in conventional metals. Since, charge carriers are confined to tight cyclotron orbits at high magnetic fields, limiting their transverse motion and causing the thermoelectric response to stop increasing with further increases in field. To understand the linear non-saturating $S_{yx}$ in ScSb, we consider the two carriers model for electronic and thermal transport. When both the electrons and holes take part in electrical transports, the $S_{yx}$ can be expressed as \cite{Delves_1964, Review_TM_AFM_2025}
\begin{equation}
S_{yx} = \frac{\sigma_{yy}^e \sigma_{yy}^h (\mu_e + \mu_h)B}{(\sigma_{yy}^e + \sigma_{yy}^h)^2 + (\sigma_{yy}^e \mu_e B -\sigma_{yy}^h \mu_h B)^2} (S_{xx}^h - S_{xx}^e)
\end{equation}
When the magnetic field is weak ($\mu B \ll 1$), the electrical conductivity dominates the denominator, causing $S_{yx}$ to grow linearly with increasing $B$. In contrast, under a strong magnetic field ($\mu B \gg 1$),  $S_{yx}$ becomes inversely proportional to $B$ if the electron and hole contributions to the Hall conductivity are significantly unbalanced. In the idealized scenario where the carrier concentrations and mobilities are equal for electrons and holes, $n_e = n_h = n$ and $\mu_e = \mu_h = \mu$, their Hall effects cancel each other out, producing the maximum possible thermomagnetic response. In this case, $S_{yx}$ can be reformulated as
\begin{equation}
    S_{yx} = \frac{S_{xx}^h - S_{xx}^e}{2}~\mu B
\end{equation}
For polycrystalline ScSb, $n_h/n_e \sim 1.1$ and $\mu_h/\mu_e \sim 1.2$, indicating a nearly ideal scenario for maximum thermomagnetic response, which leads to a linear non-saturating $S_{yx}$ even at high magnetic fields of 14 T. 

To determine the Nernst power factor ($PF_N$) and Nernst figure of merit ($z_N$), we measured the longitudinal electrical conductivity ($\sigma_{yy}$) and longitudinal thermal conductivity ($\kappa_{xx}$) as shown in Fig.\ref{Fig_PFN}(a) and \ref{Fig_PFN}(c). As the temperature increases, the $\kappa_{xx}$ initially rises, reaching a peak around 50K, and then decreases at higher temperatures. Such a peak has also been reported in several topological thermomagnetic materials exhibiting a large Nernst effect, including NbP and NbSb$_2$ \cite{NbP_poly_EEC_2018, NbSb2_poly_EES_2023}. The peak value of $\kappa_{xx}$ in polycrystalline ScSb is $\sim$ 37 W K$^{-1}$ m$^{-1}$ ( at 50 K), which is smaller than the peak value observed in polycrystalline NbP ($\sim$ 65 W K$^{-1}$ m$^{-1}$ at 85 K) and NbSb$_2$ ($\sim$ 90 W K$^{-1}$ m$^{-1}$ at 35 K) \cite{NbP_poly_EEC_2018, NbSb2_poly_EES_2023}. Under an applied magnetic field, $\kappa_{xx}$ exhibits a significant reduction, decreasing to $\sim$ 30 W K$^{-1}$ m$^{-1}$ at 14 T in polycrystalline ScSb. The measured $\kappa_{xx}$ in Fig.\ref{Fig_PFN}(c) is composed of the lattice thermal conductivity $\kappa_l$ and electronic thermal conductivity $\kappa_e$. Under magnetic field, their relationship can be expressed by the empirical formula \cite{Kxx(BT)_PRB_2002, NbSb2_NatComn_2022, NbSb2_poly_EES_2023},

\begin{equation}
\label{kxx}
\kappa_{xx}(B,T) = \kappa_l (T)+ \kappa_e (B,T) =  \kappa_l (T) + \frac{\kappa_e (0,T)}{1 + \eta B^s}
\end{equation}

where $\eta$ and $s$ are the two factors related to the thermal mobility and scattering mechanism, respectively. The increase of $B$ will suppress the contribution of carriers, which is responsible for the reduction of $\kappa_{xx}$ under high magnetic field (Fig.\ref{Fig_PFN}(c)). By using the above equation, the measured $\kappa_{xx}$ data of ScSb under different $B$ and $T$ are fitted. The fitting results are shown in Fig.\ref{Fig_PFN}(e) and table \ref{Parameters}. The electronic thermal conductivity estimated from the the Wiedemann–Franz law, $\kappa_e= L_0 \sigma_{yy} T$, is $\sim$ 12.2 W K$^{-1}$ m$^{-1}$ at 50 K, where $L_0$ = 2.44 $\times 10^{-8}$ W $\Omega$ K$^{-2}$. This value exceeds the actual $\kappa_e$ ($\sim$ 8.56 W K$^{-1}$ m$^{-1}$) obtained from fitting equation (\ref{kxx}) to the measured thermal conductivity $\kappa_{xx}$. This discrepancy indicates a violation of the Wiedemann–Franz law, possibly due to inelastic scattering of the charge carriers under a strong magnetic field, and such a violation has also been reported in NbSb$_2$ and WP$_2$ \cite{NbSb2_NatComn_2022, NbSb2_poly_EES_2023, WP2_NatComn_2018, WP2_npj_2018}.

The Nernst power factor and Nernst figure of merit are calculated using the relations $PF_N = S_{yx}^2 /\rho_{yy}$ and $z_N = S_{yx}^2/ (\rho_{yy} ~\kappa_{xx})$ as displayed in Fig.\ref{Fig_PFN}(b) and \ref{Fig_PFN}(d). In polycrystalline ScSb, $PF_N$ reaches a maximum of $\sim 240 \times 10^{-4}$ W m$^{-1}$ K$^{-2}$ at 30 K and 14 T, which is significantly higher than the value for single-crystalline ScSb ($\sim 35 \times 10^{-4}$ W m$^{-1}$ K$^{-2}$ at 12 K and 14 T) \cite{ScSb_arxiv_2025}. Furthermore, this $PF_N$ exceeds the Seebeck power factors ($PF_S$) of state-of-the-art thermoelectric materials such as Bi$_2$Te$_3$ ($PF_S$ = 42 $\times 10^{-4}$ W m$^{-1}$ K$^{-2}$) \cite{Bi2Te3_Science_2008}, CoSb$_3$ ($PF_S$ = 40 $\times 10^{-4}$ W m$^{-1}$ K$^{-2}$) \cite{CoSb3_NatMatr_2015}, PbTe ($PF_S$ = 25 $\times 10^{-4}$ W m$^{-1}$ K$^{-2}$) \cite{PbTe_Nature_2011}, and SnSe ($PF_S$ = 10 $\times 10^{-4}$ W m$^{-1}$ K$^{-2}$) \cite{SnSe_Nature_2014}. Despite ScSb’s trivial electronic band structure, the maximum $PF_N$ in polycrystalline ScSb surpasses that of several topological semimetals, including single-crystalline Cd$_3$As$_2$, ZrTe$_5$, PtSn$_4$ and TbPtBi \cite{Cd3As2_PRL_2017, ZrTe5_PRB_2021, PtSn4_Research_2020, TbPtBi_AM_2023}, as well as polycrystalline NbP and Mg$_3$Bi$_2$ \cite{NbP_poly_EEC_2018, Mg3Bi2_poly_AM_2022}.  The peak values of $PF_N$ and $z_N$ for polycrystalline ScSb are compared with those of various topological thermomagnetic single crystals and polycrystalline materials in Fig.\ref{Fig_comp}.

\section{Conclusions}
The tunability of the Nernst thermopower in ScSb primarily depends on electron-hole compensation, with the highest $S_{yx}$ achieved at perfect compensation ($n_h/n_e$ = 1). In ScSb single crystals, the Fermi level varies from crystal to crystal, resulting in either electron or hole doping, and consequently, deviations from perfect electron-hole compensation. This variability is reflected in the magnetoresistance of ScSb single crystals, which ranges from $\sim$ 500$\%$ to $\sim$ 28000$\%$ depending on the individual crystal \cite{ScSb_PRB_2018}. Controlling the level of electron or hole doping is particularly challenging in single crystals, as they are typically grown using excess Sb flux. In contrast, polycrystalline samples offer an advantage, as they are synthesized using an exact stoichiometric ratio of the constituent elements. Our polycrystalline ScSb shows a larger $S_{yx}$ ($\sim$ 128 $\mu$V/K at 30 K and 14 T) compared to its single-crystalline counterpart ($S_{yx} \sim$ 35 $\mu$V/K at 12 K and 14 T) \cite{ScSb_arxiv_2025}, likely due to improved electron-hole compensation. Polycrystalline ScSb also shows a large non-saturating magnetoresistance of $\sim 940 \%$ at 2 K and 14 T. The cubic crystal structure provides an advantage in achieving enhanced thermomagnetic and electromagnetic performance in polycrystalline form, comparable to that of single crystals. Such behavior has been observed in the cubic topological thermomagnetic material Co$_2$MnGa \cite{Co2MnGa_AEM_2024}. In contrast, other crystal structures such as tetragonal tend to exhibit a reduced average $S_{yx}$ in their polycrystalline forms. For example, while NbP single crystals show a large Nernst thermopower of 800$\mu$V/K, it drops sharply to 90$\mu$V/K in polycrystalline samples \cite{NbP_poly_EEC_2018}. Further enhancement of the Nernst effect in ScSb may be achieved by precisely tuning the electron-hole composition.

\section*{ACKNOWLEDGMENTS}
The research at Brookhaven National Laboratory was supported by the U.S. Department of Energy, Office of Basic Energy Sciences, Contract No. DE-SC0012704.
%

\end{document}